\documentclass[aps,twocolumn,groupedaddress]{revtex4}
\usepackage{epsfig}
\usepackage{dcolumn}
\usepackage{amsmath}
\begin{document}
\def\square{\kern1pt\vbox{\hrule height .6pt\hbox{\vrule width
..6pt\hskip 2.5pt
\vbox{\vskip 5pt}\hskip 2.5pt\vrule width 0.6pt}\hrule height
0.6pt}\kern1pt}
\def\btt#1{{\tt$\backslash$#1}}
\def\lmno{LaMnO$_3$}
\def\eg{$e_g$}
\def\Op{$O_p$}
\def\c{$c$}
\def\ab{$ab$}
\def\vs{$versus$}
\def\tc{$T_c$}
\def\etal{$et~al.$}
\def\ie{$i.e.$}
\def\swi{\hbox{$\sigma_{1}(\omega)$}}
\def\sw{\hbox{$\sigma(\omega)$}}
\def\cm-1{\hbox{cm$^{-1}$}}
\def\gr{\hbox{$\rho$}}
\def\w{\hbox{$\omega$}}
\def\gd{\hbox{$\delta$}}
\def\gW{\hbox{$\Omega$}}
\def\gt{\hbox{$\tau$}}
\def\gp{\hbox{$\pi$}}
\def\ga{\hbox{$\alpha$}}
\def\gl{\hbox{$\lambda$}}
\def\gu{\hbox{$\mu$}}
\def\gk{\hbox{$\kappa$}}
\def\gep{\hbox{$\epsilon$}}
\def\gs{\hbox{$\sigma$}}
\bibliographystyle{apsrev}

\preprint{}
\title{Temperature Dependence of Low-Lying Electronic Excitations of \lmno}



\author{M. A. Quijada}
\thanks{Present Address: GSFC-NASA, Code 551, Greenbelt, MD 20740}
\email[]{manuel.quijada@gsfc.nasa.gov}
\author{J. R. Simpson}
\author{H. D. Drew}
\affiliation{Materials Research Laboratory, Department of Physics, University of Maryland, College Park, Maryland 20742}
\author{J. W. Lynn}
\author{L. Vasiliu-Doloc}
\thanks{ Present address: Department of Physics, Northern Illinois University, DeKalb, IL 60115}
\affiliation{Center for Neutron Research, National Institute of Standards and Technology, Gaithersburg, Maryland 20899}
\author{Y. M. Mukovskii}
\author{S. G. Karabashev}
\affiliation{Moscow Power Engineering Institute, Moscow, Russia}

\date{\today}

\begin{abstract}
We report on the optical properties of undoped single crystal \lmno, the parent compound of the colossal magneto-resistive manganites.   Near-Normal incidence reflectance measurements are reported in the frequency range of 20$-$50,000 cm$^{-1}$ and in the temperature range 10$-$300 K. The optical conductivity, $\sigma_1(\omega)$, is derived by performing a Kramers-Kronig analysis of the reflectance data. The far-infrared spectrum of $\sigma_1(\omega)$ displays the infrared active optical phonons. We observe a shift of several of the phonon to high frequencies as the temperature is lowered through the Neel temperature of the sample ($T_N \sim 137$ K). The high-frequency $\sigma_1(\omega)$ is characterized by the onset of absorption near 1.5 eV. This energy has been identified as the threshold for optical transitions across the Jahn-Teller split $e_g$ levels. The spectral weight of this feature increases in the low-temperature state. This implies a transfer of spectral weight from the UV to the visible associated with the paramagnetic to antiferromagnetic state. We discuss the results in terms of the double exchange processes that affect the optical processes in this magnetic material.
\end{abstract}

\maketitle

\section{Introduction\newline}
Recently, there has been much attention focused on the hole-doped ferromagnetic manganite materials of the form (Ln)$_{1-x}$(A)$_{x}$MnO$_{3-\delta }$, where Ln is a lanthanide and A is an alkaline-earth element \cite{kuster4,kuster5}.  This interest is driven by  the colossal magneto-resistance (CMR) effect that is seen when doping of the alkaline-earth element is in the range of $0.2<x<0.5$ \cite{kuster6}. The CMR effect is associated with a phase transition between a paramagnetic insulator and a ferromagnetic metal at a temperature $T_c$.

Recent experimental reports suggest a very complex interplay of charge, orbital and spin order, as well as the presence of strong lattice effects in these materials \cite{kim,billinge,asamitsu,Jaime,quijadaprb}. Furthermore, it is expected that given the complex phase diagram of this class of materials, studies of the stoichiometric parent compound could give insight into the physics governing the doped version of these manganite oxides. Indeed, room-temperature optical measurements of \lmno\ have shown evidence for a static Jahn Teller (JT) distortion while in the CMR materials it is believed that the Jahn Teller effect becomes dynamic and has a strong influence on the phase transition.  Therefore studies of the parent material may allow a determination of the important parameters of the electron-phonon and electron-electron interactions required for a complete understanding of the CMR alloys.

Calculations of the optical conductivity of \lmno\ within the local-spin-density approximation suggests that the observed gap in the optical conductivity near 1$-$2 eV corresponds to a hopping transition of a hole between the JT split $ e_{g}$ bands of two adjacent Mn$^{3+}$ ions \cite{jung,hamada}. This is essentially a charge-transfer transition with a relatively weak spectral weight. Similarly, a JT feature in the optical spectrum of \lmno\ has been predicted and calculated using various models for the band structure. However, there is disagreement of the relative importance of the Coulomb repulsion energy $U$, electron-phonon, and magnetic interactions in this system. In analyzing the electronic feature in \swi\ near 1$-$2 eV, some authors argue that the effective $U$ to the low-energy physics is rather small ($U \approx 2$eV) \cite{millisprl95,millisprl,millisprb,rother}. In this view, both the effects of the JT coupling and Coulomb interactions lead to similar contributions to the gap to low-lying optical excitations in this material. Recent calculations done by Millis and Ahn using a tight-binding parameterization of the band structure, give the full conductivity tensor, with a prediction for a rather strong temperature dependence and  anisotropy in the spectrum of \swi\ for this compound \cite{ahn}. The main reason for the strong temperature dependence follows from the spin selection rules on the charge-transfer excitations and the fact that \lmno\ is a paramagnet at room temperature, while turning into an A-type antiferromagnet (AF) at low temperatures (Neel temperature $\sim 140$ K).

On the other hand and given the relatively large value for $U$ that is obtained from high-energy photoemission experiments ($U_{bare}\approx 30$ eV), other authors argue that this large $U$ rules out double occupancy that would give rise to low-lying optical transitions between neighboring Mn$^{+3}$ ions. In recent work, Allen and Perebeinos \cite{allenprl,allenprb} argue that electron-phonon interactions would lead to a phonon-assisted {\it intra}-site transition on the Mn$^{+3}$ ions. Since no hopping occurs in this model, this feature would be insensitive to the magnetic state of the Mn$^{3+}$ ions, and therefore, no change would be expected in the electronic absorption as $T$ is varied through the Neel temperature\cite{allenprl}.

In this paper, we address the issue of the temperature dependence of the optical conductivity of \lmno\ and its relevance of the calculations mentioned above by performing detailed temperature dependent measurements of \swi\ for \lmno. We obtain the spectrum of \swi\ by performing a Kramers-Kronig (KK) analysis on bulk reflectance measurements over a very wide frequency range (30-50,000 \cm-1) and for temperatures between 300 K and 10 K.
 
The results at low frequencies indicate the insulating nature of the material and show several optical phonons. Upon cooling, there is an increase in both the number and the strength of the phonon features indicating a lowering of the symmetry of the crystal structure in the AF state. There is also  a growth in spectral weight and frequency shift in some of the phonon modes as the temperature is reduced.

At higher frequencies, the spectrum of \hbox{$\sigma_{1}(\omega)$}indicates a weak absorption feature centered near 2 eV, and a large spectral feature at 4 eV. The 2 eV feature is seen to grow in oscillator strength by 25 \% at low temperatures. We attribute this effect to the different effects of double exchange on optical processes in the AF phase and the paramagnetic phase. We find that a simple model of spin alignment below $T_N$ accounts for the increase in probability of this charge-transfer transition between Hund's rule spin-split $e_g$ derived bands. These results are not consistent with the prediction of the calculations by Allen and Perebeinos \cite{allenprl}. The second and larger feature near 4 eV is mostly associated with a charge transfer transition between the O$_{2p}$ and the Mn$_d$ derived bands \cite{arima}. Although this latter feature has a strong temperature dependence, we are not able to provide a fully quantitative result, due to the uncertainties introduced by the KK analysis at this energy which is near the high energy cut off of our measurements.

\section{Experimental Details\newline}

\subsection{Sample characterization}

\begin{figure}
\includegraphics[width=3.0in,height=1.6in]{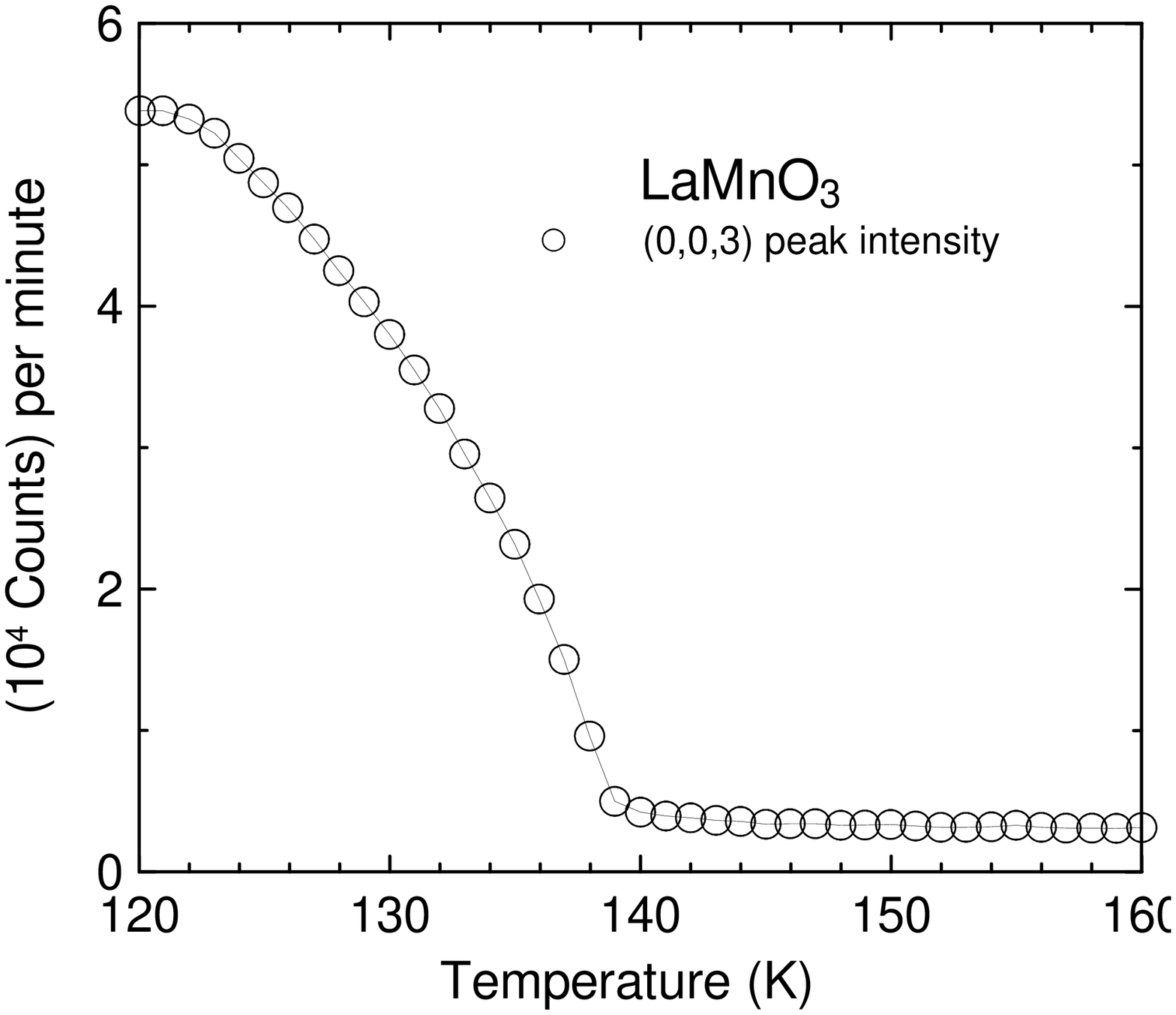}
\caption{Neutron scattering measurements on \lmno\ showing the magnetic transition near 137 K.}
\label{fig1}
\end{figure}

The samples used in this study are single crystals of the stoichiometric compound \lmno\ that were prepared by the floating-zone technique. The samples were characterized by dispersive x-ray, microwave-absorption studies\cite{lofland} and neutron scattering analysis. The neutron scattering data were taken on a BT-2 triple-axis spectrometer at the National Institute of Standards and Technology (NIST). Figure~\ref{fig1} shows the temperature dependence in the intensity of the (0,0,3) antiferromagnetic Bragg reflection measured on our \lmno\ sample.
The presence of this neutron scattering peak indicates the long-range order for the A-type antiferromagnetic alignments of the spins associated with the Mn ions. Based on the onset for detection of this peak in the data of Fig.~\ref{fig1}, we deduce a $T_N$ close to 137 K.

Further characterization was done by using X-ray diffraction measurements. These results show that the measured crystals were heavily twinned. We also found evidence of twinning by using a near-field scanning optical microscope (NSOM) probe. In this set up, a HeNe laser beam ($\lambda \sim 632.8$nm) was sent through the tip of the NSOM to illuminate the surface of the sample. We also made use of the NSOM to monitor the reflected laser beam. Changes in the reflectance as a function of polarization of the laser beam were observed as the NSOM tip was scanned through different spots on the sample. These changes were recorded as images of high- and low-brightness regions across the sample surface. These images indicated the presence of alternating long and narrow strip domains running across the sample with a typical width of $\sim 0.2 \mu$m. 

The NSOM data suggest a response in the reflectance that is highly anisotropic with respect to the crystallographic axes of the sample. Hence, the broad-frequency reflectance measurements presented in this paper represent an average of the contributions resulting from all three different crystallographic axes.

\subsection{Optical techniques}

We study the optical properties of these samples by measuring near-normal incidence reflectance over a wide range of energies (0.005-5 eV) and perform Kramers-Kronig (KK) transformations to obtain the the frequency-dependent optical conductivity \swi. Optically smooth surfaces were obtained by polishing the samples using diamond paste with particle size of 0.3 $\mu$m. Recently, Takenaka \etal\cite{takenaka} pointed out discrepancies in the reflectance of polished $\it versus$ cleaved samples of the doped versions of this compound. Given the fact they saw little difference in the reflectance of either a polished and cleaved samples of pure \lmno, we are confident the polishing of the samples of this study did not introduce significant spurious effect on the results reported here. 

Optical reflectivity was measured by using a combination of FTIR spectrometers (Bomen DA3 and Bruker IFS 113v) to cover the photon energy range of 0.005$-$5.0 eV. Temperature dependent measurements in the 10$-$300 K range were done by attaching the sample to the tip of a continuous-flow cryostat with a calibrated Si sensor mounted nearby. The accuracy in the absolute and relative changes in reflectance with temperature is estimated to be better than $\pm .5$\%. 
Analysis of the data was performed by applying Kramers-Kronig (KK) transformation to the reflectance data in order to obtain the optical properties such as \swi. The usual requirement of the KK integrals to extend the reflectance at low- and high-frequency ends was done in the following way. At low frequencies, the extension was done by keeping the reflectance constant to dc, as appropriate for insulators. 

The high frequency extrapolations were done by merging our data, which extend up to 5 eV, with published data by Jung \etal\cite{jung} which extend up to 30 eV. The range beyond 30 eV was extended using the standard power law R$\sim \omega^{-4}$, which is the free-electron behavior limit. Further details of the steps we took to maintain a temperature independent sum rule will be given in Sec.~\ref{sec:level2}.
\section{Results\protect\\} 
\subsection{Temperature Dependence of Reflectance} 
\label{sec:level2}

The results of the far infrared temperature-dependent reflectance are shown in the inset of Fig.~\ref{fig2}. These results indicate the reflectance typical of an insulating material that shows several optical phonon features. The temperature dependence in some of these phonon features is seen as a shift towards high frequency in the phonon peak positions as the temperature is lowered through $T_N$ of the sample. Further discussion on this will be done later when showing the results of \swi\ after performing the KK analysis.

In the high-frequency data shown in the main part of Fig.~\ref{fig2}. We observe two features centered at 1.5 eV and 4.2 eV respectively. Discussion on the assignment of these interband transitions will be done later. We also notice a decrease in the reflectance in the visible as the temperature is reduced down to 10 K. The temperature dependence occurs for energies higher than 1.5 eV, and it appears to be as a gradual reduction in the value of $R$ as the temperature is changed between the 300-150 K interval. Furthermore, we do not see a measurable change below about 100 K. An important observation of these data is the fact the they show temperature dependence up to highest energy of the experiment (5 eV). This is a significant observation given the fact that there is a sum rule in the reflectance that requires that the integrated difference in $R(\omega)$ from $\omega =0  \rightarrow \infty$ at two different temperatures must be equal to zero. This can be written in an equation form as:
\begin{figure}
\includegraphics[width=3.0in,height=2.3in]{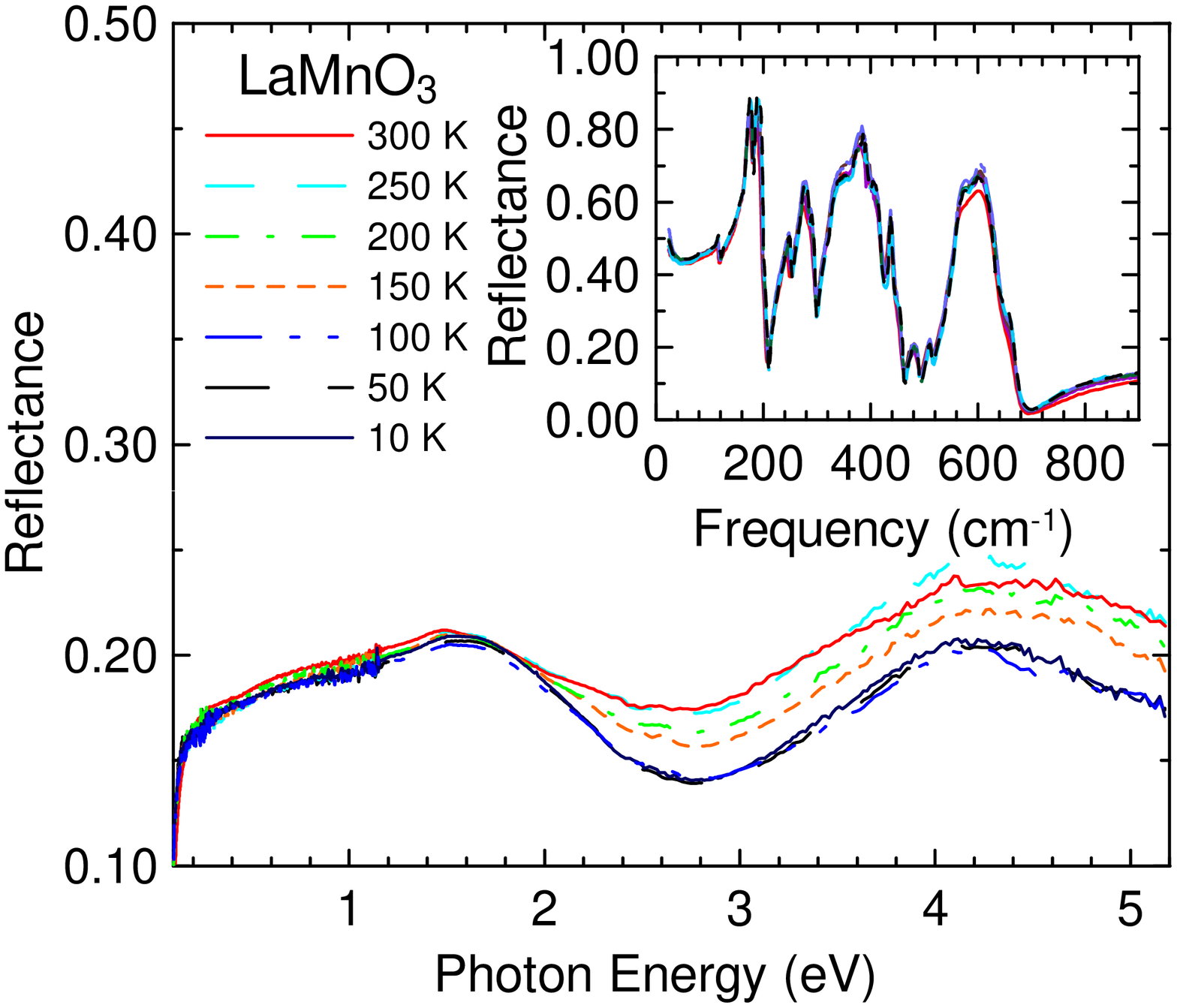}
\caption{Temperature dependence of the reflectance of the \lmno\ sample in the visible/UV range. Inset: far-IR results.}
\label{fig2}
\end{figure}

\begin{equation}
\int_0^{\infty}Ln\left[{R_1(\omega)\over{R_2(\omega)}}\right]d\omega=0,\label{zero}
\end{equation}
where $R_1(\omega)$ and $R_2(\omega)$ are the reflectance at temperatures $T_1$ and $T_2$ respectively. These results indicate that the temperature dependence  that we observe in the present data must be reversed at higher energies in order to satisfy Eq.~\ref{zero}.

The implications of these results are also relevant to the KK analysis we performed on the these data. As we mentioned earlier, the high-frequency extrapolations that are needed for the KK analysis were done by using the published room-temperature results by Jung \etal\cite{jung}, which extend up to 30 eV, and a power law $R \sim w^{-4}$ above this energy. It is important in such extensions not to introduce a step in the reflectance where the literature data connect to the experimental data. Such a step would yield false structure in the optical constants. Because there is temperature dependence in our present data all the way up to 5 eV, and in the absence of any previously published data, we performed the following extrapolation scheme in order to satisfy the reflectance sum rule of Eq.~\ref{zero}. The scaling factor used in appending the 300 K data from Jung \etal\ to our low-temperature results was increased in the 5-10 eV range and then joined smoothly to the unchanged data above 10 eV \footnote{Changing this energy to 20 eV did not have a significant impact on the KK results presented in this work.}. For each individual temperature below 300 K, the scaling factor was adjusted so that the result of Eq.~\ref{zero} were temperature independent above 10-12 eV. The implication of this is that there is a transfer of spectral weight between the range where we measured $R$ and higher energies.  Furthermore, this implies a temperature dependence of some interband electronic transitions at energies higher than 5 eV. It should be pointed out that varying this extrapolation procedure did not make any significant effect on the results of the KK analysis below about 3 eV.

We could also write a partial sum rule function related to the carrier density, $N_{eff}(\omega)$, participating in the optical transition at a frequency $\omega$. This is defined as:
\begin{equation}
N_{eff}(\omega)={2m^*V_{cell}\over{4\pi e^2}}\int_0^{\omega}\swi d\omega,\label{one}
\end{equation}
where \swi\ is the optical conductivity, $m^*$ is the electronic mass and $V_{cell}$ is the unit cell volume. When the integral of Eq.~\ref{one} is carried out from zero up to $\omega = \infty$ it gives a result that represents the total number of carriers in the system, and it also must be temperature invariant. This is related to the result expressed in Eq.~\ref{zero}.

\subsection{Far-infrared Phonon Spectrum \protect\\} 

\begin{figure}
\includegraphics[width=3in,height=2.2in]{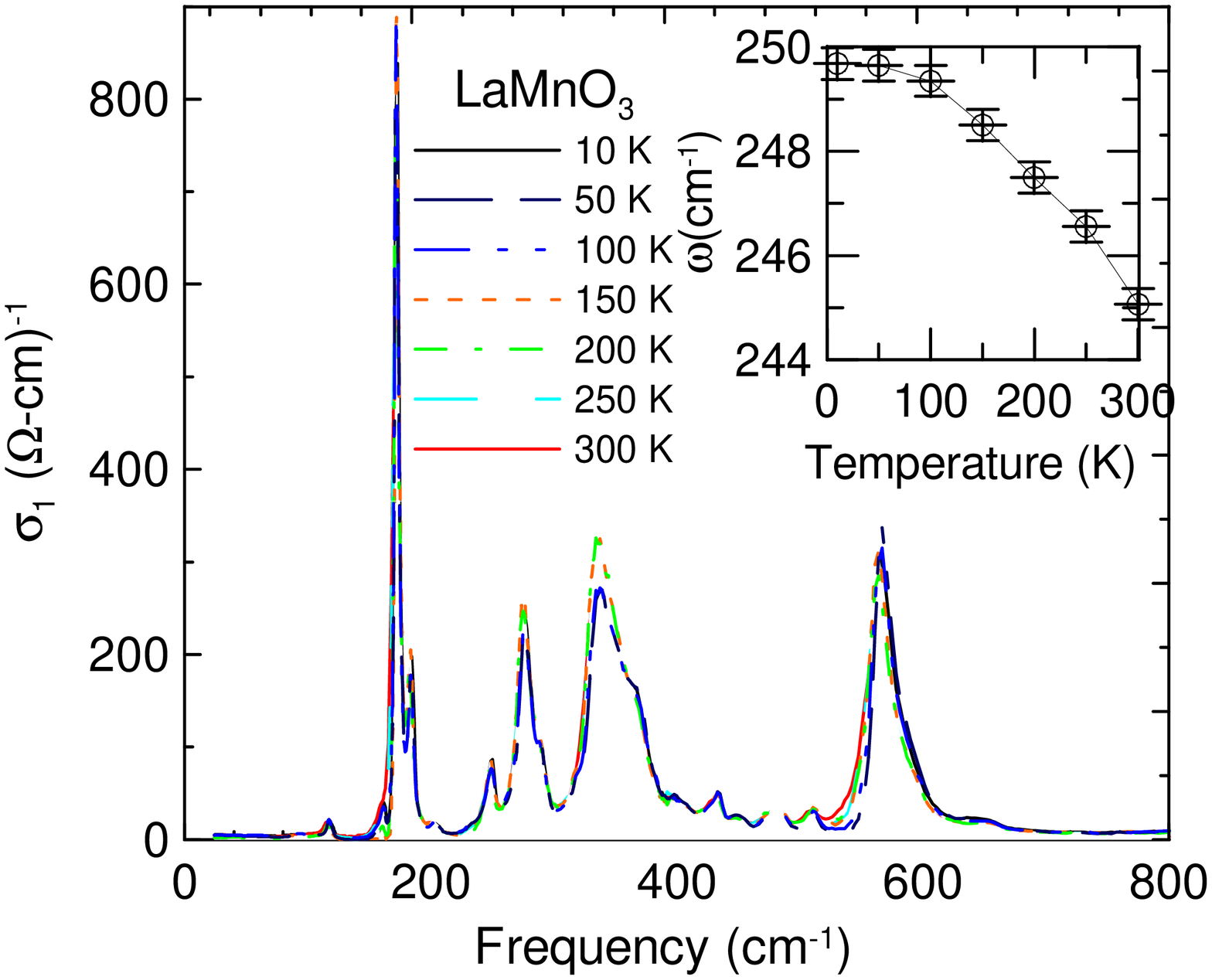}
\caption{ Optical conductivity, \swi, showing the phonon spectrum from KK analysis of data in Fig. 2.}
\label{fig3} 
\end{figure}

Fig.~\ref{fig3} shows the result of the real part of the optical conductivity, \swi\ obtained from a KK analysis of the reflectance data at far-infrared frequencies. We observe several peaks in the spectrum of \swi\ that correspond to transverse optical (TO) phonon vibrations of the atoms in the system. We summarized the results of a fit using Lorentz oscillators of the most prominent phonon features at 300 K and 10 K in Table~\ref{table1}. First of all, we notice all the modes show a shift of the peak position to higher frequencies at low temperatures. This is highlighted in the inset of Fig.~\ref{fig3}, where we show a representative temperature shift of the phonon peak near 250-243 \cm-1.  This shift could be due to an overall thermal contraction of the unit cell. Another observation is that, as usual, the phonon features become narrower and sharper at low temperatures. This behavior is due to the fact that in general, phonon lifetimes are dominated by thermal fluctuations. It is worth noting the unusually narrow linewidth of the mode near 170 \cm-1\ ($\gamma \sim 3.2 \cm-1$) at 10 K. Because of its frequency position, this mode most likely involves vibrations of the heavier La atom in this compound.
\begin{table}
\caption{Parameters from Lorentz fits to the major phonon features of \lmno\ at 300 K and 10 K. $\omega_{j}$ is the TO frequency, $\gamma_{j}$ is the linewidth, and $\omega_{jp}$ represents the oscillator strength all in units of ${\rm cm}^{-1}$.}
\label{table1}
\begin{ruledtabular}
\begin{tabular}{lccccccccc}
300 K & & & & & & & & & \\ 
Osc. \# &1&2&3&4&5&6&7&8& \\ 
$\omega_{j}$&115& 170 & 182 & 245 & 274 & 335 & 359 & 560 &\\ 
$\gamma_{j}$ &4.9& 6.6 & 6.3 & 8.5 & 13.4 & 23.0 & 43.2 & 20.0& \\ 
$\omega_{jp}$&69& 430 & 175 & 139 & 342 & 550 & 514 & 457 &\\ 
 && & & & & & & & \\ 
10 K & & & & & & & & \\ 
Osc. \# &1&2&3&4&5&6&7&8&9 \\ 
$\omega_{j}$&118& 173 & 184 & 250 & 276 & 336 & 363 & 566&645 \\ 
$\gamma_{j}$ &4.5& 3.2 & 5.0 & 8.3 & 9.8 & 20.6 & 34.0 & 16.4&44.7 \\ 
$\omega_{jp}$&70& 430 & 229 & 187 & 361 & 564 & 517 & 507&172 \\ 
\end{tabular}
\end{ruledtabular}
\end{table}
A final observation is the fact that some of the phonon modes are more clearly resolved at low temperatures. This is reflected on the fact that the oscillator strength parameter, $\omega_{jp}$, is larger at $T=10$K, as the fit results show in Table~\ref{table1}.  This is particularly true for the modes near 184, 250, and 645 \cm-1. In fact, the latter mode is so hard to discern at room temperature that we only show the Lorentz parameters at 10 K. Such increase in the number of phonon modes suggests a decrease of the symmetry at low temperatures. This is consistent with Raman measurements\cite{romero} that show a reduction in the intensity of some Raman-active phonon modes at high temperatures. This is also consistent with the material gradually becoming  more cubic in the paramagnetic state. Finally, the growth in $\omega_{jp}$ may also suggest that some of the bonds involved in these vibrations become less covalent (and more ionic) at low temperatures.

Recently, Paolone \etal\cite{paolone} have discussed assignment of nearly all of the infrared active modes in \lmno. Since our spectra are very similar to the ones presented by these authors, we will not discuss this issue further here. Instead, we will devote the rest of the paper to presenting the results of the temperature dependence in the electronic absorptions at higher energies.

\subsection{Temperature dependence of electronic contribution\protect\\} 

\begin{figure}
\includegraphics[width=3in,height=2.2in]{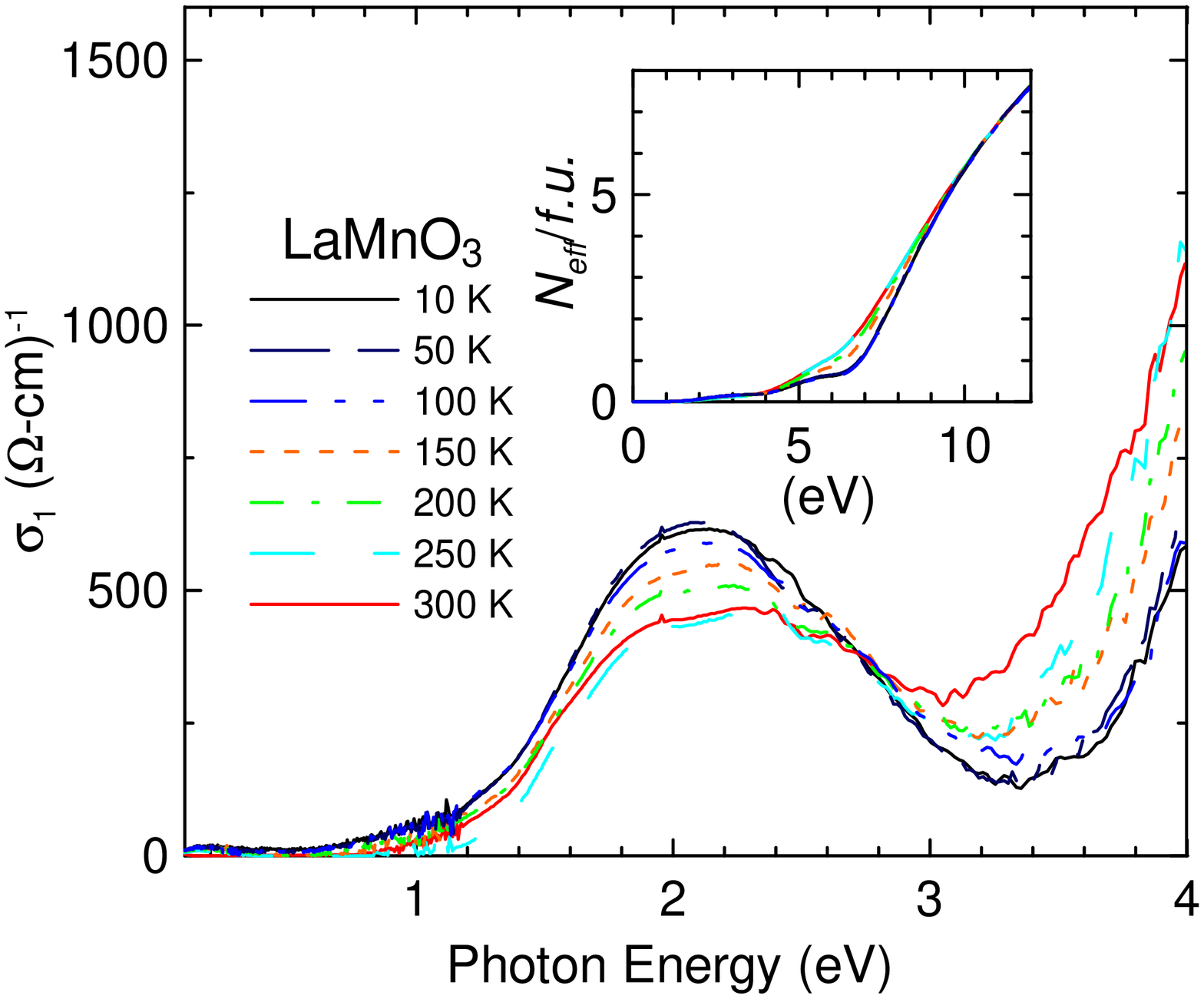}
\caption{Temperature dependence in the optical conductivity showing an increase in spectral weight of the 2 eV feature at low temperatures, followed by a decrease of spectral weight at higher energies. Inset: Sum rule analysis of \swi\ from Eq.2}
\label{fig4} 
\end{figure}

We now turn to the electronic part of the spectrum of the optical conductivity. Fig.~\ref{fig4} displays the results of \swi\ in the 1$-$4 eV range for this material at several temperatures. The results at 300 K show two main structures; a lower peak centered at 2.0 eV followed by second larger peak centered around 4$-$5 eV. These energies are in good agreement with previous room-temperature conductivity measurements for this compound \cite{arima,jung,takenaka}. In those earlier works the 2.0 eV peak has been interpreted as the transition between the JT-split \eg$-$\eg\ levels within the parallel spin manifold. The energy position of transitions to the spin reversed states ($J_H$) has been suggested to be in the 3$-$4 eV range. This means that this transition may overlap with the charge transfer transition between the Mn$-3d$ \eg\ and the oxygen$-2p$ \Op\ levels. The energy position for the latter one is interpreted to be in 4$-$5 eV range. This degeneracy is one of the reasons why there has been some controversy regarding the correct assignment for the energy position of $J_H$\cite{okimoto}.

A remarkable observation, which to the best of our knowledge has so far not been reported for this material, is the temperature dependence that we show in fig.~\ref{fig4}. First, we observe that at 300 K, the low-energy peak is centered near $\sim $ 2 eV, it has an intensity of 470 $\Omega^{-1}\cm-1$ with a width of 1.5 eV.  This result is in overall agrement with the conductivity spectrum reported earlier \cite{jung,takenaka}. At 10 K, we notice a sharpening of the 2.0 eV feature (width $\sim 1.3$ eV) and an increase in intensity (\swi$_{peak} \sim$ 600 $\Omega^{-1} \cm-1$). Also, there appears to be no significant shift in the energy position of the peak. 

At the same time, there is a marked reduction in the intensity of the conductivity structure above 4 eV.  We are not able to fully quantify the changes in temperature of the 4 eV energy feature due to the uncertainties in the KK extrapolations.  However, the results for the 2 eV peak are quite robust.  Overall, however, the results indicate a gradual redistribution of spectral weight as the temperature is reduced, \ie\ the oscillator strength of the 2 eV feature increases at low temperature by as much as 25 \% at 10 K. This is followed by a reduction of spectral weight for the transitions at higher energies, related to the exchange transition $J_H$ and the charge transfer transition between the Mn$-3d$ and the oxygen$-2p$. We emphasize that much of the temperature dependence we are reporting here occurs between 300 K and 150 K. In the next section, we will present an interpretation of these results in term of the A-type antiferromagnetic phase transition that occurs in this system near 140 K. We will also compare these results with theoretical calculations of the optical conductivity of this compound.

\section{Discussion\protect\\} 

\begin{figure}
\includegraphics[width=3in,height=2.2in]{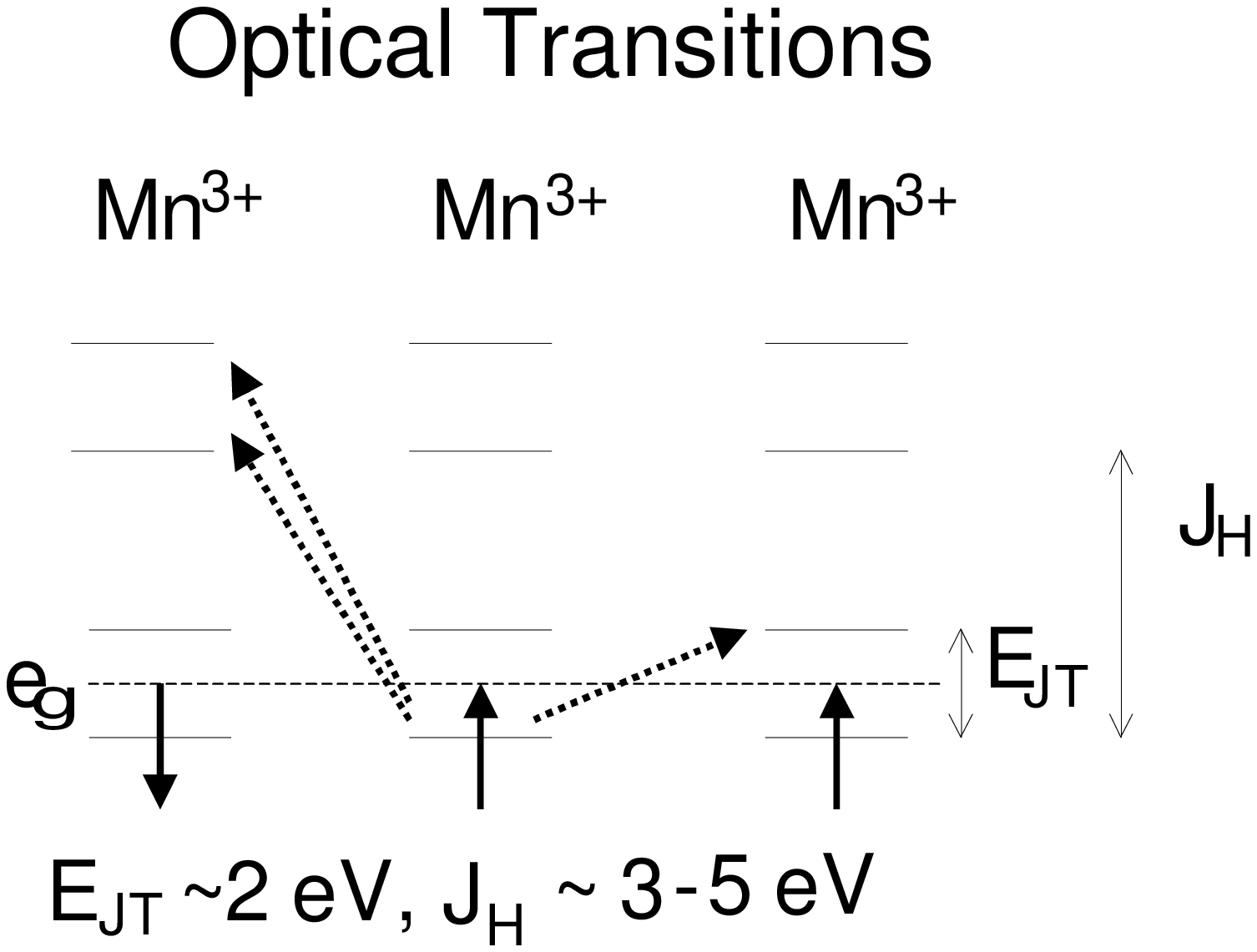}
\caption{Diagram showing the optical processes among the low-lying states in \lmno.}
\label{fig5} 
\end{figure}

The origin of the observed optical transitions in \lmno\ has been previously discussed by Solovyev \etal\cite{hamada} and Terakura \etal\cite{terakura} using the Local-Spin-Density (LDA) approximation. These results, which do not take into consideration any effects due to changes in temperature, model the gap feature near 2.0 eV for this compound, as an optical transition between JT-split \eg\ levels.  In interpreting the results presented here, we would make an extension of their basic picture that the peak at 2.0 eV is the result of a hopping transition between JT-split \eg\ levels by including the effects of the spin selection rules on this transition. In this context, the temperature dependence seen in fig.~\ref{fig4}, could be explained, at least qualitatively, by the diagram of fig.~\ref{fig5}. In this diagram, we show a localized energy picture of nearest-neighbor Mn$^{3+}$ ions, and the possible optical transitions. Hence, we denote the lowest electronic transition as $E_{JT}$, and it is the hopping transition between JT-split \eg\ levels of adjacent ions. Next, we denote a second energy, $J_H$, related to the Hund's coupling energy that involves transitions to a higher energy spin-flip state. The relative changes as a function of temperature could be explained in the following way. In the paramagnetic state, above $T_N$, there is equal probability for spins of neighboring ions to be aligned parallel or antiparallel.  However, below T$_N$ the spins order in a A-type antiferromagnetic state, in which there are ferromagnetic planes which are stacked antiferromagnetically. This gives a 67\% probability that the neighboring spins are aligned $versus$  a 33\% probability for anti-parallel alignment.  This suggest that relatively, the hopping transition between JT-split \eg\ levels has a higher probability below T$_N$. This should also be followed by a corresponding reduction of the transitions to the Hund's split levels $J_H$ above the JT-split transitions. This is exactly what we observe in the results of fig.~\ref{fig4}. Below T$_N$, the strength of the 2.0 eV peak increases by about 25 \%, while we observe a reduction of oscillator strength at higher energy.  Although the data clearly show this trend, a more quantitative comparison is not possible for the 4 eV feature due to increase uncertainties of the KK analysis close to the high-energy cut off of our experiment.

While the quantitative success of this simple model is no doubt accidental it certainly supports the basic underlying picture of the microscopic optical processes operating in the manganites.   However, the model ignores the anisotropy of the hopping matrix elements among the eg orbitals.  A quantitative comparison with a more realistic calculation would be highly desirable.

Recent calculations by Ahn and Millis\cite{ahn} offer an opportunity to compare the experimental results presented here with more elaborate calculations. In this work, the authors performed calculations based on  a model Hamiltonian using a tight binding parameterization of the band structure, along with a mean-field treatment of Hund, electron-electron, and electron-lattice couplings. Among the key results of these calculations are the prediction of a strong anisotropy and temperature dependence in the total spectral weight of the 2.0 eV feature in the optical spectrum of \lmno. The results of these calculations can be most revealingly compared to the results in Fig.~\ref{fig4} by means of the spectral weight or, equivalently, the electronic kinetic energy ($K(\omega)$) associated with the 2 eV feature.  The kinetic energy is defined in Ref.~[\onlinecite{ahn}].  Hence, it follows from the partial sum-rule expression of $N_{eff}(\omega)$ given in Eq.\ (\ref{one}) that 
\begin{equation}
K(\omega)={a_{0}\over{mV_{cell}}}N_{eff}(\omega),\label{two}
\end{equation} where $a_0$ represents the unit cell lattice parameter. The integral in Eq.\ (\ref{one}) is carried out up to $\omega \sim$ 2.0 eV, to account only for the optical transitions of interest in this case. We show in Table~\ref{table2} the results of these calculations both at 300 K and 10 K. When the results of Table~\ref{table2} are compared with theoretical estimates of the average kinetic energy ($K_{ave}$) from Reference [\cite{ahn}], we find that both the experiment and the theory predict an increase in the kinetic energy value $K$ for the JT peak at $T=0$ (In the experiment $T=10$K). In the theory, this temperature dependence results from the double-exchange-driven correlation between spin-dependent hopping amplitudes (as is also suggested in the diagram shown in fig.~\ref{fig5}). Such behavior in the theory becomes even more robust when sensible values are chosen for the coupling constant $\lambda$, the Hund's coupling energy ($J_HS_c$) and the Coulomb $U$. 

We should point out that the theory overestimates the total spectral value for $K_{ave}$ for the 2.0 eV peak. This is mostly attributed to Coulomb interactions. The numbers reported in Table~\ref{table2} are assuming $U=0$, along with $\lambda \simeq 1.38$ eV/$\AA$, and $J_HS_c \simeq 2.0$. More realistic assumptions for $U$ substantially reduce the total kinetic energy value for $K_{ave}$ for the 2 eV feature. The authors in Ref.~[\onlinecite{ahn}] deduce a value for the Coulomb repulsion energy of $U \simeq 1.6$ . This has the effect of making the results of the theory and experiment to be in closer agreement. Another consideration in the theory is the value of the coupling constant parameter $\lambda$ and the exchange energy $J_H$. Some estimates have the energy for $J_H$ to be in the 3-4 eV range\cite{quijadaprb}. Unfortunately, the close proximity of this transition with the charge transfer transition between the O $2p$ and Mn $d$ makes it hard to separate the exact temperature dependence of the $J_H$ transition in our data.

As pointed out in Sec.~\ref{sec:level2} and can be seen in fig.~\ref{fig4}, the temperature dependence of the optical conductivity occurs mostly above \tc.  This observation is initially surprising in terms of the interpretation presented here of the optical processes operating in this system.  In this interpretation the temperature dependence of the optical conductivity of the 2 eV feature is related to the temperature dependence of the spin alignment of neighboring Mn ions it may be expected that it would follow the temperature dependence of the magnetization of the sample.  However, the magnetization is a measure of the long range order of the spins while the optical charge transfer process depends on the local spin order.  Therefore, the optical conductivity gives a probe of the short range order of the spin system.  The experiment shows, in fact, that short range spin order sets in well above \tc\ in this system.  These results suggest that optical measurements could be used to make a quantitative study of short range spin order in this system.     

\begin{table}
\caption{Total spectral weight of optical transition at 2.0 eV . The theoretical values are obtained from Ref.~[\protect\onlinecite{ahn}], for $T$ = 0 and $U$ = 0.} 
\label{table2}
\begin{ruledtabular}
\begin{tabular}{lcc}
 & Experiment & Theory \\ 
$K_{ave}$(300 K)& 0.066 eV & 0.086 eV \\
$K_{ave}$(10 K) &  0.082 eV  & 0.095 eV \\ 
\end{tabular}
\end{ruledtabular}
\end{table}

\section{Conclusions\newline}
In conclusion, we have studied the temperature dependence in the optical conductivity of \lmno, the parent compound of the colossal magneto resistance manganites. We find a temperature dependence in the optical absorption near 2 eV, that is consistent with the magnetic transition that occurs in this system below 140 K. A simple picture of counting the spin alignment of nearest-neighbors gives a qualitative explanation for the changes in oscillator strength that are observed. These results are in contradiction to the predictions by Allen \etal\ that this peak could be explained in terms of on site optical transitions of a self-trapped exciton resulting from electron-phonon interactions.  Indeed, our results demonstrate that the dominant contribution to the optical spectral weight of the conductivity peak at 2.0 eV is the charge transfer hopping between nearest neighbor manganese ions.  The comparison of these optical measurements with recent model calculations provides estimations of some of the key parameters that are important for the manganites. 

\begin{acknowledgments}

We would like to thank S. E. Lofland for performing the x-ray analysis on these samples, and R. Decca for performing the NSOM measurements. We would also like to thank A. Millis for useful discussions. This work was supported in part by the NSF-MRSEC grant \# DMR-96-32521 and DMR-9705482.
\end{acknowledgments}

\bibliography{lmoD}

\begin{thebibliography}{10}
\expandafter\ifx\csname bibnamefont\endcsname\relax
  \def\bibnamefont#1{#1}\fi
\expandafter\ifx\csname bibfnamefont\endcsname\relax
  \def\bibfnamefont#1{#1}\fi
\expandafter\ifx\csname url\endcsname\relax
  \def\url#1{\texttt{#1}}\fi
\expandafter\ifx\csname urlprefix\endcsname\relax\def\urlprefix{URL }\fi
\providecommand{\bibinfo}[2]{#2}
\providecommand{\eprint}[2][]{\url{#2}}

\bibitem{kuster4}
\bibinfo{author}{\bibfnamefont{S.}~\bibnamefont{Jin}},
  \bibinfo{author}{\bibfnamefont{T.~H.} \bibnamefont{Tiefel}},
  \bibinfo{author}{\bibfnamefont{M.}~\bibnamefont{McCormack}},
  \bibinfo{author}{\bibfnamefont{R.~A.} \bibnamefont{Fastnacht}},
  \bibinfo{author}{\bibfnamefont{R.}~\bibnamefont{Ramesh}}, \bibnamefont{and}
  \bibinfo{author}{\bibfnamefont{L.~H.} \bibnamefont{Chen}},
  \bibinfo{journal}{Science} \textbf{\bibinfo{volume}{264}},
  \bibinfo{pages}{413} (\bibinfo{year}{1994}).

\bibitem{kuster5}
\bibinfo{author}{\bibfnamefont{M.}~\bibnamefont{McCormack}},
  \bibinfo{author}{\bibfnamefont{S.}~\bibnamefont{Jin}},
  \bibinfo{author}{\bibfnamefont{T.~H.} \bibnamefont{Tiefel}},
  \bibinfo{author}{\bibfnamefont{R.~M.} \bibnamefont{Fleming}},
  \bibinfo{author}{\bibfnamefont{J.~M.} \bibnamefont{Phillips}},
  \bibnamefont{and} \bibinfo{author}{\bibfnamefont{R.}~\bibnamefont{Ramesh}},
  \bibinfo{journal}{Appl. Phys. Lett.} \textbf{\bibinfo{volume}{64}},
  \bibinfo{pages}{3407} (\bibinfo{year}{1994}).

\bibitem{kuster6}
\bibinfo{author}{\bibfnamefont{H.~L.} \bibnamefont{Ju}},
  \bibinfo{author}{\bibfnamefont{C.}~\bibnamefont{Kwon}},
  \bibinfo{author}{\bibfnamefont{Q.}~\bibnamefont{Li}},
  \bibinfo{author}{\bibfnamefont{R.~L.} \bibnamefont{Greene}},
  \bibnamefont{and}
  \bibinfo{author}{\bibfnamefont{T.}~\bibnamefont{Venkatesan}},
  \bibinfo{journal}{Appl. Phys. Lett.} \textbf{\bibinfo{volume}{65}},
  \bibinfo{pages}{2109} (\bibinfo{year}{1994}).

\bibitem{kim}
\bibinfo{author}{\bibfnamefont{K.}~\bibnamefont{Kim}},
  \bibinfo{author}{\bibfnamefont{J.}~\bibnamefont{Gu}},
  \bibinfo{author}{\bibfnamefont{E.}~\bibnamefont{Choi}},
  \bibinfo{author}{\bibfnamefont{G.}~\bibnamefont{Park}}, \bibnamefont{and}
  \bibinfo{author}{\bibfnamefont{T.}~\bibnamefont{Noh}},
  \bibinfo{journal}{Phys. Rev. Lett.} \textbf{\bibinfo{volume}{77}},
  \bibinfo{pages}{1877} (\bibinfo{year}{1996}).

\bibitem{billinge}
\bibinfo{author}{\bibfnamefont{S.~J.~L.} \bibnamefont{Billinge}},
  \bibinfo{author}{\bibfnamefont{R.~G.} \bibnamefont{Difrancesco}},
  \bibinfo{author}{\bibfnamefont{G.~H.} \bibnamefont{Kwei}},
  \bibinfo{author}{\bibfnamefont{J.~J.} \bibnamefont{Neumeier}},
  \bibnamefont{and} \bibinfo{author}{\bibfnamefont{J.~D.}
  \bibnamefont{Thompson}}, \bibinfo{journal}{Phys. Rev. Lett.}
  \textbf{\bibinfo{volume}{77}}, \bibinfo{pages}{715} (\bibinfo{year}{1996}).

\bibitem{asamitsu}
\bibinfo{author}{\bibfnamefont{A.}~\bibnamefont{Asamitsu}},
  \bibinfo{author}{\bibfnamefont{Y.}~\bibnamefont{Moritomo}},
  \bibinfo{author}{\bibfnamefont{Y.}~\bibnamefont{Tomioka}},
  \bibinfo{author}{\bibfnamefont{T.}~\bibnamefont{Arima}}, \bibnamefont{and}
  \bibinfo{author}{\bibfnamefont{Y.}~\bibnamefont{Tokura}},
  \bibinfo{journal}{Nature} \textbf{\bibinfo{volume}{373}},
  \bibinfo{pages}{407} (\bibinfo{year}{1995}).

\bibitem{Jaime}
\bibinfo{author}{\bibfnamefont{M.}~\bibnamefont{Jaime}},
  \bibinfo{author}{\bibfnamefont{M.~B.} \bibnamefont{Salamon}},
  \bibinfo{author}{\bibfnamefont{K.}~\bibnamefont{Pettit}},
  \bibinfo{author}{\bibfnamefont{M.~R.~R.} \bibnamefont{E.Treece}},
  \bibinfo{author}{\bibfnamefont{J.~S.} \bibnamefont{Horwitz}},
  \bibnamefont{and} \bibinfo{author}{\bibfnamefont{D.~B.}
  \bibnamefont{Chrisey}}, \bibinfo{journal}{Appl. Phys. Lett,}
  \textbf{\bibinfo{volume}{68}}, \bibinfo{pages}{1576} (\bibinfo{year}{1996}).

\bibitem{quijadaprb}
\bibinfo{author}{\bibfnamefont{M.}~\bibnamefont{Quijada}},
  \bibinfo{author}{\bibfnamefont{J.}~\bibnamefont{Cerner}},
  \bibinfo{author}{\bibfnamefont{J.~R.} \bibnamefont{Simpson}},
  \bibinfo{author}{\bibfnamefont{H.}~\bibnamefont{Drew}},
  \bibinfo{author}{\bibfnamefont{K.~H.} \bibnamefont{Ahn}},
  \bibinfo{author}{\bibfnamefont{A.~J.} \bibnamefont{Millis}},
  \bibinfo{author}{\bibfnamefont{R.}~\bibnamefont{Shreekala}},
  \bibinfo{author}{\bibfnamefont{R.}~\bibnamefont{Ramesh}},
  \bibinfo{author}{\bibfnamefont{C.}~\bibnamefont{Kwon}},
  \bibinfo{author}{\bibfnamefont{M.}~\bibnamefont{Rajeswari}},
  \bibnamefont{and}
  \bibinfo{author}{\bibfnamefont{T.}~\bibnamefont{Venkatesan}},
  \bibinfo{journal}{Phys. Rev. Lett.} \textbf{\bibinfo{volume}{77}},
  \bibinfo{pages}{2081} (\bibinfo{year}{1996}).

\bibitem{jung}
\bibinfo{author}{\bibfnamefont{J.~H.} \bibnamefont{Jung}},
  \bibinfo{author}{\bibfnamefont{K.~H.} \bibnamefont{Kim}},
  \bibinfo{author}{\bibfnamefont{T.~W.} \bibnamefont{Noh}},
  \bibinfo{author}{\bibfnamefont{E.~J.} \bibnamefont{Choi}}, \bibnamefont{and}
  \bibinfo{author}{\bibfnamefont{J.}~\bibnamefont{Yu}}, \bibinfo{journal}{Phys.
  Rev. B} \textbf{\bibinfo{volume}{57}}, \bibinfo{pages}{11043}
  (\bibinfo{year}{1998}).

\bibitem{hamada}
\bibinfo{author}{\bibfnamefont{I.}~\bibnamefont{Solovyev}},
  \bibinfo{author}{\bibfnamefont{N.}~\bibnamefont{Hamada}}, \bibnamefont{and}
  \bibinfo{author}{\bibfnamefont{K.}~\bibnamefont{Terakura}},
  \bibinfo{journal}{Phys. Rev. B} \textbf{\bibinfo{volume}{53}},
  \bibinfo{pages}{7158} (\bibinfo{year}{1996}).

\bibitem{millisprl95}
\bibinfo{author}{\bibfnamefont{A.~J.} \bibnamefont{Millis}},
  \bibinfo{author}{\bibfnamefont{P.~B.} \bibnamefont{Littlewood}},
  \bibnamefont{and} \bibinfo{author}{\bibfnamefont{B.~I.}
  \bibnamefont{Shraiman}}, \bibinfo{journal}{Phys. Rev. Lett.}
  \textbf{\bibinfo{volume}{74}}, \bibinfo{pages}{5144} (\bibinfo{year}{1995}).

\bibitem{millisprl}
\bibinfo{author}{\bibfnamefont{A.~J.} \bibnamefont{Millis}},
  \bibinfo{author}{\bibfnamefont{R.}~\bibnamefont{Mueller}}, \bibnamefont{and}
  \bibinfo{author}{\bibfnamefont{B.~I.} \bibnamefont{Shraiman}},
  \bibinfo{journal}{Phys. Rev. Lett.} \textbf{\bibinfo{volume}{77}},
  \bibinfo{pages}{175} (\bibinfo{year}{1996}).

\bibitem{millisprb}
\bibinfo{author}{\bibfnamefont{A.~J.} \bibnamefont{Millis}},
  \bibinfo{author}{\bibfnamefont{R.}~\bibnamefont{Mueller}}, \bibnamefont{and}
  \bibinfo{author}{\bibfnamefont{B.~I.} \bibnamefont{Shraiman}},
  \bibinfo{journal}{Phys. Rev. B} \textbf{\bibinfo{volume}{54}},
  \bibinfo{pages}{5405} (\bibinfo{year}{1996}).

\bibitem{rother}
\bibinfo{author}{\bibfnamefont{H.}~\bibnamefont{Roder}},
  \bibinfo{author}{\bibfnamefont{J.}~\bibnamefont{Zang}}, \bibnamefont{and}
  \bibinfo{author}{\bibfnamefont{A.~R.} \bibnamefont{Bishop}},
  \bibinfo{journal}{Phys. Rev. Lett.} \textbf{\bibinfo{volume}{76}},
  \bibinfo{pages}{1356} (\bibinfo{year}{1996}).

\bibitem{ahn}
\bibinfo{author}{\bibfnamefont{K.~H.} \bibnamefont{Ahn}} \bibnamefont{and}
  \bibinfo{author}{\bibfnamefont{A.~J.} \bibnamefont{Millis}},
  \bibinfo{journal}{Phys. Rev. B} \textbf{\bibinfo{volume}{61}},
  \bibinfo{pages}{13545} (\bibinfo{year}{2000}).

\bibitem{allenprl}
\bibinfo{author}{\bibfnamefont{P.~B.} \bibnamefont{Allen}} \bibnamefont{and}
  \bibinfo{author}{\bibfnamefont{V.}~\bibnamefont{Perebeinos}},
  \bibinfo{journal}{Phys. Rev. Lett.} \textbf{\bibinfo{volume}{83}},
  \bibinfo{pages}{4828} (\bibinfo{year}{1999}).

\bibitem{allenprb}
\bibinfo{author}{\bibfnamefont{P.~B.} \bibnamefont{Allen}} \bibnamefont{and}
  \bibinfo{author}{\bibfnamefont{V.}~\bibnamefont{Perebeinos}},
  \bibinfo{journal}{Phys. Rev. B} \textbf{\bibinfo{volume}{60}},
  \bibinfo{pages}{10747} (\bibinfo{year}{1999}).

\bibitem{arima}
\bibinfo{author}{\bibfnamefont{T.}~\bibnamefont{Arima}},
  \bibinfo{author}{\bibfnamefont{Y.}~\bibnamefont{Tokura}}, \bibnamefont{and}
  \bibinfo{author}{\bibfnamefont{J.}~\bibnamefont{Torrance}},
  \bibinfo{journal}{Phys. Rev. B} \textbf{\bibinfo{volume}{48}},
  \bibinfo{pages}{17006} (\bibinfo{year}{1993}).

\bibitem{lofland}
\bibinfo{author}{\bibfnamefont{S.~E.} \bibnamefont{Lofland}},
  \bibinfo{author}{\bibfnamefont{S.~M.} \bibnamefont{Bhagat}},
  \bibinfo{author}{\bibfnamefont{H.~C.} \bibnamefont{Xiong}},
  \bibinfo{author}{\bibfnamefont{T.}~\bibnamefont{Venkatesan}},
  \bibinfo{author}{\bibfnamefont{R.~L.} \bibnamefont{Greene}},
  \bibnamefont{and} \bibinfo{author}{\bibfnamefont{S.}~\bibnamefont{Tyagi}},
  \bibinfo{journal}{J. Appl. Phys.} \textbf{\bibinfo{volume}{79}},
  \bibinfo{pages}{5166} (\bibinfo{year}{1996}).

\bibitem{takenaka}
\bibinfo{author}{\bibfnamefont{K.}~\bibnamefont{Takenaka}},
  \bibinfo{author}{\bibfnamefont{K.}~\bibnamefont{Iida}},
  \bibinfo{author}{\bibfnamefont{Y.}~\bibnamefont{Sawaki}},
  \bibinfo{author}{\bibfnamefont{S.}~\bibnamefont{Sugai}},
  \bibinfo{author}{\bibfnamefont{Y.}~\bibnamefont{Moritomo}}, \bibnamefont{and}
  \bibinfo{author}{\bibfnamefont{A.}~\bibnamefont{Nakamura}},
  \bibinfo{journal}{J. Phys. Soc. Jpn.} \textbf{\bibinfo{volume}{68}},
  \bibinfo{pages}{1828} (\bibinfo{year}{1999}).

\bibitem{romero}
\bibinfo{author}{\bibfnamefont{V.~B.} \bibnamefont{Podobedov}},
  \bibinfo{author}{\bibfnamefont{A.}~\bibnamefont{Weber}},
  \bibinfo{author}{\bibfnamefont{D.~B.} \bibnamefont{Romero}},
  \bibinfo{author}{\bibfnamefont{J.~P.} \bibnamefont{Rice}}, \bibnamefont{and}
  \bibinfo{author}{\bibfnamefont{H.~D.} \bibnamefont{Drew}},
  \bibinfo{journal}{Phys. Rev. B} \textbf{\bibinfo{volume}{58}},
  \bibinfo{pages}{43} (\bibinfo{year}{1998}).

\bibitem{paolone}
\bibinfo{author}{\bibfnamefont{A.}~\bibnamefont{Paolone}},
  \bibinfo{author}{\bibfnamefont{P.}~\bibnamefont{Roy}},
  \bibinfo{author}{\bibfnamefont{A.}~\bibnamefont{Pimenov}},
  \bibinfo{author}{\bibfnamefont{A.}~\bibnamefont{Loidl}},
  \bibinfo{author}{\bibfnamefont{O.~K.} \bibnamefont{mel'nikov}},
  \bibnamefont{and} \bibinfo{author}{\bibfnamefont{A.~Y.}
  \bibnamefont{Shapiro}}, \bibinfo{journal}{Phys. Rev. B}
  \textbf{\bibinfo{volume}{61}}, \bibinfo{pages}{11255} (\bibinfo{year}{2000}).

\bibitem{okimoto}
\bibinfo{author}{\bibfnamefont{Y.}~\bibnamefont{Okimoto}},
  \bibinfo{author}{\bibfnamefont{T.}~\bibnamefont{Katsufuji}},
  \bibinfo{author}{\bibfnamefont{T.}~\bibnamefont{Ishikawa}},
  \bibinfo{author}{\bibfnamefont{T.}~\bibnamefont{Arima}}, \bibnamefont{and}
  \bibinfo{author}{\bibnamefont{Y.Tokura}}, \bibinfo{journal}{Phys. Rev. B}
  \textbf{\bibinfo{volume}{55}}, \bibinfo{pages}{4206} (\bibinfo{year}{1997}).

\bibitem{terakura}
\bibinfo{author}{\bibfnamefont{K.}~\bibnamefont{Terakura}},
  \bibinfo{author}{\bibfnamefont{I.}~\bibnamefont{Solovyev}}, \bibnamefont{and}
  \bibinfo{author}{\bibfnamefont{S.}~\bibnamefont{Sawada}},
  \emph{\bibinfo{title}{Colossal Magnetoresistive Oxides}}
  (\bibinfo{publisher}{Gordon and Breach}, \bibinfo{address}{Tokyo},
  \bibinfo{year}{1999}).

\end{thebibliography}
%
%

\end{document}